\documentclass{IEEEtran}

\ifCLASSINFOpdf
  \usepackage[pdftex]{graphicx}
  \graphicspath{./imgs/}
  \DeclareGraphicsExtensions{.pdf,.jpeg,.png}
\else
  \usepackage[dvips]{graphicx}
  \graphicspath{./imgs/}
  \DeclareGraphicsExtensions{.eps}
\fi

\usepackage{epstopdf}
\usepackage{tikz}
\usepackage{url}
\usepackage[cmex10]{amsmath}

\usepackage{mathtools} 
\usepackage{color}
\usepackage{cleveref}

\usepackage[ruled,vlined]{algorithm2e}

\usepackage{cite}

\usepackage[export]{adjustbox}

\ifCLASSOPTIONcompsoc
  \usepackage[caption=false,font=normalsize,labelfont=sf,textfont=sf,justification=centering]{subfig}
\else
  \usepackage[caption=false,font=footnotesize,justification=centering]{subfig}
\fi
\usepackage{stfloats}
\renewcommand*{\Re}{\operatorname{Re}} 
\renewcommand*{\Im}{\operatorname{Im}} 

\DeclareMathOperator*{\argmax}{argmax} 
\DeclareMathOperator*{\argmin}{argmin} 

\usepackage{algpseudocode}
\usepackage{amsfonts}
\usepackage{float}
\usepackage{multirow}

\usepackage[colorinlistoftodos]{todonotes}

\algnewcommand\algorithmicinput{\textbf{Input:}}
\algnewcommand\Input{\item[\algorithmicinput]}
\algnewcommand\algorithmicoutput{\textbf{Output:}}
\algnewcommand\Output{\item[\algorithmicoutput]}

\begin{document}
%
\title{Forced Oscillation Identification and Filtering from Multi-Channel Time-Frequency Representation}
%
%
%

\author{Pablo Gill Estevez, Pablo Marchi, Francisco Messina, and Cecilia Galarza 
\thanks{P. Gill Estevez, P. Marchi, F. Messina, and C. G. Galarza work with the School of Engineering, Universidad de Buenos Aires and the CSC-CONICET, Argentina. (e-mail: pgill@fi.uba.ar, pmarchi@fi.uba.ar, fmessina@fi.uba.ar, cgalar@fi.uba.ar)}}%
\maketitle

\begin{abstract}
Location of non-stationary forced oscillation (FO) sources can be a challenging task, especially under resonance condition with natural system modes. In this case, the magnitudes of the oscillations could be greater in distant places from the source and the oscillation spreads over a large region of the power system. \textcolor{black}{Detection, }\textcolor{black}{frequency identification and filtering of FO oscillatory components constitutes an initial and critical step for the application of} \textcolor{black}{oscillation source location (OSL) methods.} \textcolor{black}{Specifically, this step has a major impact on the performance of the} \textcolor{black}{OSL} \textcolor{black}{ method, such as the Dissipating Energy Flow (DEF) method.}
In this paper we develop a systematic methodology for \textcolor{black}{detection,} identification and filtering of non-stationary FO based on multi-channel time-frequency (TF) representation (TFR). 
We compare three TF approaches applied together with the DEF method: short-time Fourier transform (STFT), STFT-based synchrosqueezing transform (FSST) and second order FSST (FSST2). We have used simulated signals and real world PMU data to shown that the proposed method provides a systematic framework for the identification and filtering of power systems non-stationary forced oscillations.

\end{abstract}


%
\IEEEpeerreviewmaketitle

\section{Introduction}
%
%
%
%

\IEEEPARstart{F}{orced} oscillations (FOs) are determined by disturbances that drive the system, unlike modal oscillations which mainly depend on the dynamic characteristics of the system \cite{Follum2016, Follum2017}. FOs can occur due to different causes, such as equipment failure, inadequate control designs, and abnormal generator operating conditions \cite{Nerc2017}. The most efficient way for mitigating sustained oscillations is to locate the source and to disconnect it from the network \cite{Chen2013}. 
Among the most effective \textcolor{black}{oscillation source location (OSL) methods}, the Dissipating Energy Flow (DEF) method \cite{Chen2013} has shown a good performance \cite{Dan2018}, and it was recently adopted by the Independent System Operator - New England (ISO-NE) \cite{Maslennikov2020,MASLENNIKOV201755}.


\textcolor{black}{The first step in many}
\textcolor{black}{OSL methods consists of detection,}
\textcolor{black}{identification and filtering of oscillatory components of the signals. It is an important step that can have a major impact on the performance of the OSL method.} 
\textcolor{black}{Detection of FO requires identifying the existence of oscillatory signatures in the measurements and also defining the thresholds to be used to differentiate oscillatory signatures from ambient changes \cite{Nerc2021}.}
\textcolor{black}{ After a forced oscillation has been detected, its parameters (such as the FO frequency) can be identified. Then, it is usually required to filter the signals to reconstruct the oscillatory components in the time domain for the application of an OSL method.} \textcolor{black}{Generally, the problems of FO detection, identification and filtering are treated separately.} 

\textcolor{black}{FO detection methods can be split into categories, based on \cite{NASPI2017}: an increase in signal energy \cite{donelly2015,Follum2016}, an increase in coherence \cite{Zhou2013}, and identification of sustained oscillations \cite{Liu2008}.}
\textcolor{black}{In the first category, \cite{donelly2015} proposed RMS energy filters to estimate the total RMS energy of a signal in defined frequency bands, while \cite{Follum2016} designed a detection threshold with associated probabilities of detection and false alarm for a periodic FO with stationary frequency. In the second category, \cite{Zhou2013} proposed examining the spectral coherence between PMU measurements from different areas of a power system. During ambient conditions, the spectral coherence between measured data tends to be small, while significant correlation at a specific frequency is indicative of a forced oscillation present in both sets of measurements \cite{NASPI2017}. Methods in the third category use algorithms initially intended to detect poorly damped modal oscillations to detect sustained forced oscillations. Subsequent analysis would be needed to distinguish whether the underlying cause of the low damping is related to poor damping of natural modes or from the presence of external forced oscillations\cite{Nerc2021}.}
\textcolor{black}{From these detection methods it is not possible to reconstruct directly the FO components in the time domain and these usually have to be combined with some specific filtering technique \cite{NASPI2017}.}

\begin{figure}[t]
\centering
\vspace*{-2mm}
\includegraphics[width = 1\columnwidth]{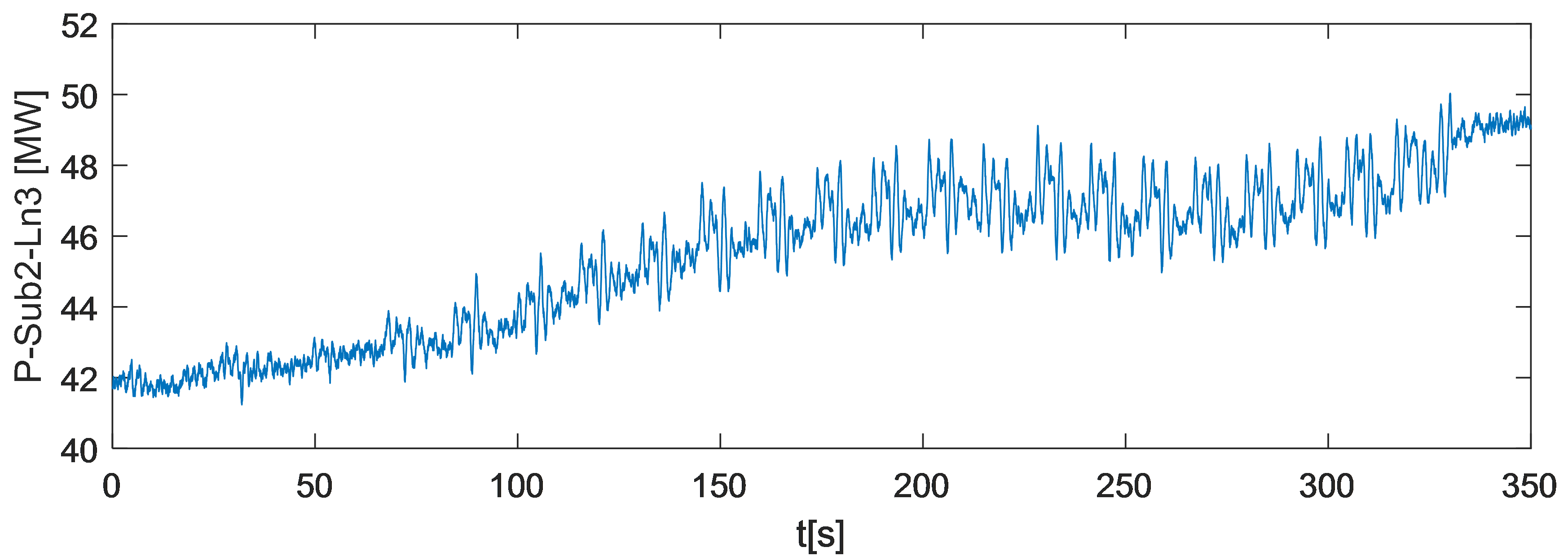}
\vspace*{-5mm}
\caption{Active Power Flow during October 3, 2017 event in ISO-NE \cite{Maslennikov2018}.}
\vspace*{-2mm}
\label{fig:Pintro}
\end{figure}

\begin{figure*}[h]
\centering
\setlength\fboxsep{0pt}
\setlength\fboxrule{0.25pt}
\includegraphics[width=2\columnwidth]{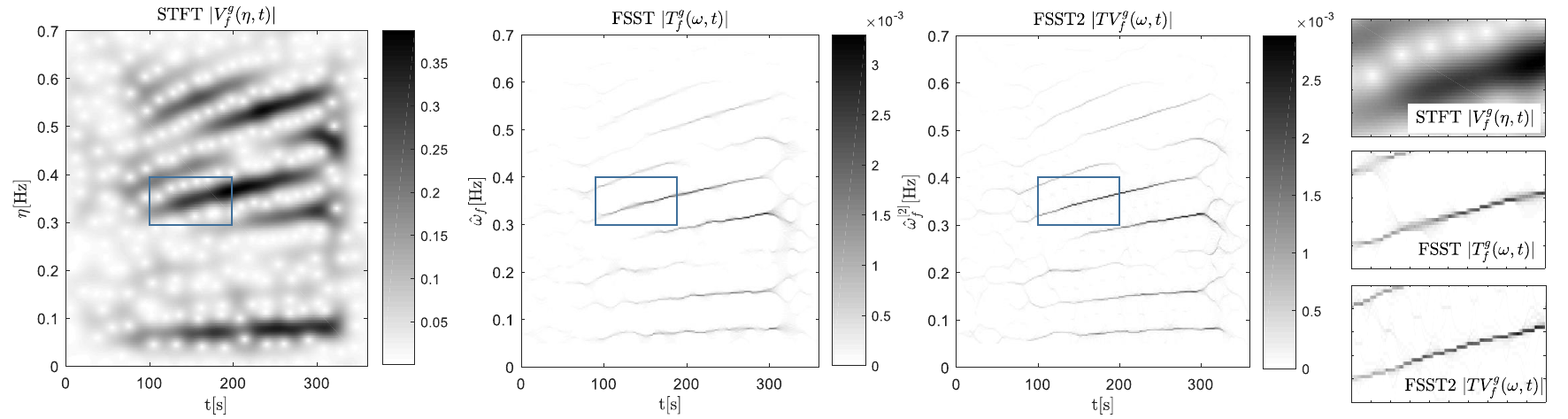}
\caption{STFT, FSST and FSST2 of active power flow P-Sub2-Ln3 during October 3, 2017 event in ISO-NE \cite{Maslennikov2018}.}
\label{fig:ejemplo_inicial_2}
\end{figure*}

\textcolor{black}{Most of the methods currently used to identify and filter FO assume that the FO frequency presents a stationary behavior over a certain time window. In this way, discrete Fourier transform (DFT) is commonly used to identify the FO frequency and then the filtering stage is typically done with bandpass filters \cite{MASLENNIKOV201755,Kirihara2019,NASPI2017}. However, the behavior of FO frequency could be considerably non-stationary, and in this case the performance of the identification and filtering methods that assume that the FO frequency is stationary over a time window might not be good enough. As an example of non-stationary FO in the real world,} Fig. \ref{fig:Pintro} shows active power flows through transmission line Sub2-Ln3 of October 3, 2017 event in ISO-NE \cite{Maslennikov2018}. Non-sinusoidal and non-stationary nature of FOs can be identified. 

Non-stationary oscillating signals are typically represented as multicomponent signals (MCSs) by a superposition of amplitude- and frequency-modulated (AM-FM) components. \textcolor{black}{Over the last few decades, non-stationary signal processing methods have attracted a lot of interest owing to their relevance and applicability to a large class of real world signals in different engineering problems \cite{Rehman2019}.} \textcolor{black}{The first proposed method was the empirical mode decomposition (EMD) algorithm \cite{Huang1998}, in the late nineties. EMD decomposes input data into its oscillatory components through a recursive "sifting" process that makes use of signal extrema \cite{Huang1998}. 
Despite its advantages, due to the lack of a theoretical framework, there is no guarantee of obtaining a good performance using the EMD decomposition \cite{Rehman2019}.
Another class of data-driven methods that aim to extract EMD-like decomposition, but with an adequate theoretical foundation, include  variational mode decomposition (VMD) \cite{Dragomiretskiy2014} and synchrosqueezing transform (SST) \cite{DAUBECHIES2011243} based on time-frequency (TF) analysis.}

\textcolor{black}{VMD decomposes a MCS into modulated oscillations exhibiting limited bandwidth across a center frequency. That is achieved by formulating and solving a convex optimization problem that minimizes the sum of the bandwidths of all oscillatory components. VMD decomposes the signal in a similar way to a band-pass filter bank, with the advantage that its parameters are systematically adjusted based on the data \cite{Dragomiretskiy2014,Rehman2019,ARRIETA2019}.}

\textcolor{black}{On the other hand, }TF analysis extracts the temporal and spectral information from the signal simultaneously, providing a suitable framework for studying non-stationary signals. The information of the different components of a MCS in the TF plane spreads around curves commonly called ridges. \textcolor{black}{FO oscillatory components are identified on the TF plane by applying a ridge identification algorithm on the coefficients of the TF representation (TFR) and then extracted and transformed into the time domain. In this case, it is not necessary to associate a frequency band to each oscillatory component of the signal (as in VMD). Thus, different oscillatory components of the signal present at different time instants over a specific frequency band could be identified separately.}

The short-time Fourier transform (STFT) is probably one of the best known TFRs. Furthermore, SST was originally introduced in the context of Continuous Wavelet Transform (CWT) \cite{DAUBECHIES2011243} and then extended to STFT \cite{Oberlin2014}, referred as STFT-based SST (FSST). FSST reassigns the information of STFT, sharpening the representation in the TF plane. 
The applicability of SST is restricted to MCSs made of slightly modulated harmonic modes. In order to deal with stronger amplitude and frequency modulations, an extension based on linear chirp approximation called second-order STFT-based SST (FSST2) was introduced in \cite{Oberlin2015,Behera2018}. As an example, Fig. \ref{fig:ejemplo_inicial_2} shows STFT, FSST and FSST2 of the signal in Fig. \ref{fig:Pintro}. We can observe multiple regions with variable frequency and amplitude in the TF plane, centered around the ridges (in this case ridges represent signal harmonics). The details in the figure show how FSST improves the concentration around the ridges over STFT and,  FSST2 improves it over FSST. 

\textcolor{black}{The main contribution of this paper is to develop a systematic methodology that jointly performs detection, identification and filtering from multi-channel TFR to retrieve the oscillatory components of non-stationary power system FOs.}
\textcolor{black}{Through the proposed methodology, it is intended to cover the limitations of the detection methods (which do not include the reconstruction of the signal components in the time domain) and to improve the performance on non-stationary FO of the identification and filtering methods reported in the literature (which assume that the behavior of the FO frequency is stationary).}
\textcolor{black}{The multi-channel terminology implies that measurements from different locations of the power system are used to generate a single TFR. We improve and extend in a more general approach the oscillation identification methodology for non-stationary forced oscillations of our more recent work \cite{Gill2020}.}

\textcolor{black}{The proposed methodology does not represent an OSL method in itself, but rather it is a systematic signal pre-processing methodology specially designed for the analysis of non-stationary signals, used as a previous step to the subsequent application of an OSL method. 
In this work, the methodology is combined with the DEF method, but it could be applied in combination with some other method that requires decomposing the data into its oscillatory components.}
In particular, we compared the performance of the method applied to simulated data and real world PMU data using three different TFRs (STFT, FSST and FSST2). 
\textcolor{black}{Additionally, we show the advantages of the proposed approach compared to other detection, identification and filtering techniques.}

\section{Mathematical Background}

\subsection{Multi-component Signal Model}

Non-stationary oscillatory data $f(t)$ is represented by a MCS as a superposition of H oscillatory components as:
\begin{equation}
f(t)=\sum_{h=1}^{H} f_h(t)+r(t),
\label{eq:sumamodos}
\end{equation}
where each oscillatory component $f_h(t)=A_h(t)e^{j2 \pi \, \phi_h (t)}$ has a time-varying amplitude $A_h(t)$ and instantaneous frequency (IF) $\phi'_h(t)$. The signal $r(t)$ represents noise plus low frequency trend signal \cite{THAKUR2013}. We further assume that the modes are separated in frequency with resolution $\Delta$: $\forall h\leq H-1, \phi'_{h+1}(t)-\phi'_h(t)>2\Delta$. In the TF plane, the different components are associated with curves called ridges, which are denoted as $(\varphi'_h)_{h=1,...,H}$, and these are estimates of the IFs $(\phi'_h)_{h=1,...,H}$ of the components.

\subsection{Short-time Fourier transform (STFT)}
We denote by $\hat{f}$ the Fourier transform of function $f$ with the following normalization:
\begin{equation}
\hat{f}(\eta) = \int_{\mathbb{R}} f(x) e^{-2j\pi \eta x
} dx
\label{eq:ft_def}
\end{equation}

The STFT is a local version of the Fourier transform obtained by means of a sliding window $g$ \cite{Oberlin2014}:
\begin{equation}
V_f^g(t,\eta) = \int_{\mathbb{R}} f(u) g(u-t) e^{-2j \pi \eta (u - t)} du
\label{eq:stft1}
\end{equation}

\textcolor{black}{In this work we consider a Gaussian window $g(t)=e^{-\pi t^2/\sigma^2}$. The standard deviation is $std_{{g}}=\sigma/(\sqrt{2\pi})$ and the standard deviation of its Fourier transform $\hat{g}$ is $std_{\hat{g}}=1/(\sqrt{2\pi}\sigma)$, both characterized by the parameter $\sigma$.}

We consider that the signal $\mathbf{f}$ is a discrete time sequence of length $L$ and time duration T, such that $\mathbf{f}[n] = f(\frac{nT}{L} )$, for $n=0,...,L-1$, and $\mathbf{g}[n]= g(\frac{nT}{L} )$ for the Gaussian window, which is truncated to be supported on ${n=-M,...,M}$ such that $2M+1\leq N$, where $N$ is the number of frequency bins. The sampling frequency is $F_s=L/T$. In that context, the discrete STFT of $\mathbf{f}$ is defined by \cite{Laurent2020}
\begin{align}
\mathbf{V}_f^g[m,k]&=\frac{1}{N}\sum_{n \in \mathbb{Z}} \mathbf{f}[n]\mathbf{g}[n-m]e^{-2j\pi\frac{k(n-m)}{N}}\\
&=\frac{1}{N}\sum_{n=-M}^{M} \mathbf{f}[m+n]\mathbf{g}[n]e^{-2j\pi\frac{kn}{N}}
\label{eq:STFTdiscrete}
\end{align}
with $k\in\{0,...,N-1\}$ and $m\in\{0,...,L-1\}$. The index $k$ corresponds to frequency $\frac{kL}{NT}$, and the index $m$ to time $\frac{mT}{L}$. \textcolor{black}{The discrete STFT can be computed for each $m$ through a discrete Fourier transform (DFT) of length N}. The reconstruction formula is
\begin{equation}
\mathbf{f}[m]=\frac{1}{g(0)}\sum_{k=0}^{N-1} \mathbf{V}_f^g[m,k]
\label{eq:reconSTFTdisc}
\end{equation}

If we assume slow variations in $A_h(t)$ and on the IF $\phi'_h(t)$ of a MCS, we can write the following approximation in the vicinity of a fixed time $t_0$:
\begin{align}
f(t) \approx \sum_{h=1}^{H} A_h(t_0) e^{j2 \pi [\phi_h (t_0)+\phi'_h(t_0)(t-t_0)]}
\end{align}

The corresponding approximation for the STFT then results (changing $t_0$ by a generic $t$):
\begin{align}
V_f^g\left(t, \eta \right) \approx \sum_{h=1}^{H} f_h(t) \hat{g}(\eta-\phi'_h(t))
\end{align}

The representation of this multicomponent signal by $V_f^g\left(t, \eta \right)$ in the TF plane shows that the peaks are concentrated around ridges defined by $\eta=\phi'_h(t)$. The frequency width around each ridge is related to the frequency bandwidth of $\hat{g}$, which can be estimated as $3std_{\hat{g}}$.

\subsection{STFT-based synchrosqueezing transform}


Starting from STFT, the FSST moves the coefficients $V_f^g(t,\eta)$ according to the map  $( t, \eta) \mapsto ( t, \hat{\omega_f} ( t, \eta ))$, where $\hat{\omega_f} ( t,\eta)$ is a local estimation of the IF defined as \cite{Oberlin2014}:
\begin{equation}
\hat{\omega}_f(t,\eta)=\frac{1}{2\pi}\partial_{t}\arg(V_f^g(t,\eta))=\eta+ \Im{\left\lbrace \frac{V_f^{g'}(t,\eta)}{2\pi V_f^{g}(t,\eta)} \right\rbrace}
\label{eq:RVw}
\end{equation}
FSST coefficients are given by \cite{Oberlin2014}:
\begin{equation}
T_f^g(t,\omega) = \frac{1}{g(0)} \int_{0}^{\infty} V_f^g\left(t, \eta \right) \, \delta\left[\omega - \hat{\omega}_f \left( t, \eta \right) \right] \, d\eta 
\label{eq:fsst}
\end{equation}
where $\delta$ denotes the Dirac distribution. 

Calculating the FSST with (\ref{eq:fsst}) sharpens the information relative to components in the TF plane around the ridges. Ridges $\varphi'_h(t)$ associated to the $h$th component are estimated with the application of a specific algorithm, described later. Each component $f_h$ can be recovered by integrating $T_s^g\left(f,t \right)$ around a small frequency band $d$ around the curve $\varphi'_h(t)$ \cite{Oberlin2014}:
\begin{equation}
f_h^{FSST}(t) = \int_{|\omega-\varphi'_h(t)|<d} T_f^g(t,\omega) \, d\omega
\label{eq:ifsst}
\end{equation}
The discrete-time version of $T_f^g(t,\omega)$ is denoted by $\mathbf{T}_f^g[m,k]$. 

\subsection{Second-order STFT-based synchrosqueezing transforms}

FSST assumes $\phi''(t)$ is negligible but in many situations the signal exhibits high frequency modulation \cite{Behera2018} and the applicability of FSST could be restricted. For this cases, an extension based on linear chirp approximation was introduced, called second-order STFT-based synchrosqueezing transform (FSST2) \cite{Oberlin2015}\cite{Behera2018}. It uses a more accurate IF estimate than $\hat{\omega_f}$, named as second-order instantaneous frequency estimator of $f$ defined by \cite{Oberlin2015}

\begingroup
\small
\begin{align}
\hat{\omega}_f^{[2]} (t,\eta)=    \begin{cases}
      \hat{\omega}_f(t,\eta)+\hat{q}_f(\eta,t)(t-\hat{\tau}_f(t,\eta)) & \text{if }  \partial_t\hat{\tau}_f(t,\eta) \neq 0\\
      \hat{\omega}_f(t,\eta) & \text{otherwise}\\
    \end{cases} 
\label{eq:Pf}
\end{align}
\endgroup

\textcolor{black}{where the group delay $\hat{\tau}_f(t,\eta)$ is calculated by \cite{Oberlin2015}:}
\begin{equation}
\textcolor{black}{
\hat{\tau}_f(t,\eta)=t+\Re{\left\lbrace \frac{V_f^{tg}(t,\eta)}{V_f^{g}(t,\eta)} \right\rbrace}
}
\label{eq:RVt}
\end{equation}

and the modulation operator $\hat{q}_f(\eta,t)$ is computed by \cite{Oberlin2015}:
 \begin{equation}
\hat{q}_f(\eta,t) = \Re{\left\lbrace \frac{1}{2i\pi}\frac{V_f^{g''}V_f^{g}-(V_f^{g'})^2}{(V_f^{g})^2+V_f^{tg}V_f^{g'}-V_f^{tg'}V_f^{g}} \right\rbrace} 
\label{eq:qsst2}
\end{equation}

To enlighten the notation we write $V_f^{g}$ instead of $V_f^{g}(t,\eta)$. If $h(t)=A(t)e^{2i\pi \phi(t)}$ is a linear chirp with $\phi(t)$ and $\log(A(t))$ quadratic functions, then $\hat{q}_h=\phi''$ and $\hat{\omega}_h^{[2]} (t,\eta)=\phi'(t)$ \cite{Meignen2015}.

FSST2 coefficients are defined by \cite{Oberlin2015}
 \begin{equation}
TV_f^g(t,\omega) = \frac{1}{g(0)} \int_{0}^{\infty} V_f^g\left(t, \eta \right) \, \delta\left[\omega - \hat{\omega}^{[2]}_f \left( t, \eta \right) \right] \, d\eta 
\label{eq:fsst2}
\end{equation}

A procedure analogous to that used with FSST can be used for ridge identification and component reconstruction from FSST2:
\begin{equation}
f_h^{FSST2}(t) = \int_{|\omega-\varphi'_h(t)|<d} TV_f^g(t,\omega) \, d\omega
\label{eq:ifsst2}
\end{equation}

Its discrete-time counterpart is denoted by $\mathbf{TV}_f^g[m,k]$.

\textcolor{black}{For the FSST2 computation it is necessary to perform 5 STFT operations (associated with $V_f^{g}$, $V_f^{g'}$, $V_f^{g''}$, $V_f^{tg}$, $V_f^{tg'}$), while for the FSST it is only necessary to perform 2 STFT operations ($V_f^{g}$, $V_f^{g'}$). Hence, adding to this the other operations for the calculation of coefficients, FSST2 computation time results approximately twice the time required for FSST. As an example, using a computer with an Intel core i5 3.2GHz and 8GB RAM, the average computation time for STFT is 0.2 s, for FSST is 0.4 s and for FSST2 is 0.8 seconds, approximately, for a 90 s time series of PMU data and a time step of 100 ms.}

\subsection{Dissipating Energy Flow Method}

The first steps of the DEF method are \cite{MASLENNIKOV201755}:

\begin{itemize}
\item \textit{PMU data pre-processing}. PMU angles for voltages and currents should be unwrapped. Replace missing PMU data (NaN) and outliers with interpolated data. Extract low frequency trend of the signals.
\item \textit{Frequency identification} of the sustained oscillation.
\item \textit{Filtering the component of interest} around the identified frequency applied to the variables of interest for DEF calculation (i.e. active and reactive power, voltage magnitude, angle, and frequency).
\end{itemize}

The flow of dissipating energy, for specific filtered components in a branch $ij$ from bus $i$ to bus $j$, is expressed by integrating over the system trajectory as follows \cite{MASLENNIKOV201755}:

\begin{equation}
W_h^{ij} \approx \int P_h^{ij} \, d\theta_h^i + \int Q_h^{ij} \frac{dv_h^i}{v^i},
\end{equation}
where $P^{ij}$ and $Q^{ji}$ are the active and reactive power flows in branch $ij$, $\theta^i$ is the voltage angle, and $v^i$ is voltage magnitude of from bus. Subindex $h$ indicates that the filtered magnitudes corresponding to $h$ oscillatory component. The integration limits are determined from the instant when sustained oscillations have significant magnitude.

The value and sign of the rate of change of $W_h^{ij}$ have a physical interpretation as the amount and direction of the dissipating energy flow. It indicates the direction of the source location relative to the branch $ij$. Positive rate of change of $W_h^{ij}$ means the source is located behind bus $i$, and a negative value means the source is located behind bus $j$ or branch $ij$ is the source. Finding a source of dissipating energy is equivalent to the finding of a source of negative damping\cite{Maslennikov2020}. For discrete PMU signals, a discrete-time approximation has the form:
\begin{align}
W_h^{ij}[m] &= W_h^{ij}[m-1] +P_h^{ij}[m-1] \, \left(\theta_h^i[m] - \theta_h^i[m-1] \right) \nonumber \\
&+ \frac{Q_h^{ij}[m-1]}{v^i[m-1]} \left( v_h^i[m] - v_h^i[m-1] \right) 
\label{eq:flow_disip_energy}
\end{align}
where $m$ represents the time instant. 

\textcolor{black}{The conventional application of the DEF method is intended for signals with stationary behavior in frequency \cite{MASLENNIKOV201755}. Discrete Fourier transform (DFT) is used to identify $f_{si}$, the frequency of interest of the sustained oscillation. Then, band-pass filtering around the identified frequency is applied to the variables of interest for DEF calculation (i.e. active and reactive power, voltage magnitude, angle, and frequency). Filter design specifications are: Butterworth filter with the pass frequencies $f_{pi}=(1 \pm e)f_{si}$ where $e$=0.05; cutoff frequencies $f_{ci}=(1 \pm 2e)f_{si}$; 1 dB of ripple allowed and 10–15 dB attenuation at both sides of the passband \cite{MASLENNIKOV201755}.}

\section{Proposed Methodology}
\begin{figure}[h]
\centering
\vspace*{-2mm}
\includegraphics[width = 0.9\columnwidth]{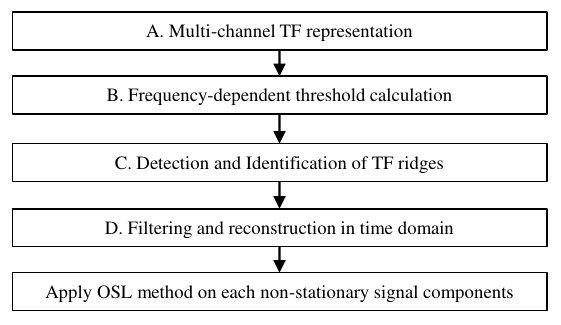}
\caption{Scheme of proposed methodology}
\vspace*{-2mm}
\label{fig:dia_ppal}
\end{figure}

Fig. \ref{fig:dia_ppal} shows the scheme of proposed methodology. The methodology could be applied to any invertible TF representation. $TF_f(t,\omega)$ and $\mathbf{TF}_f[m,k]$ are used to generically name the coefficients for the continuous and the discrete time version of the TF representations of signal $f$.  

\subsection{Multi-channel TF representation}


It is assumed that PMU measurements are available from $B$ branches at different locations in the system. In order to capture the relevant oscillatory components for DEF calculation, we define a global TF representation of the system:

\begin{equation}
\mathbf{MTF}[m,k] = \sqrt{\sum_{i=1}^{B}|\mathbf{TF_{P^i}}[m,k]|^2+|\mathbf{TF_{Q^i}}[m,k]|^2}
\label{eq:MTF}
\end{equation}

where $P^i$ and $Q^i$ are the measurements of the active power flows and the reactive power flows of branches $i=1,...,B$. $P^i$ presents good observability for components of the electromechanical range, and is not as influenced by frequencies lower than 0.1 Hz as is the bus voltage angle \cite{MASLENNIKOV201755}. $Q^i$ allows to identify components which could be less observable in $P^i$.

\subsection{Frequency-Dependent Threshold Calculation}
From measurements prior to the appearance of FO, a spectrum-dependent threshold $\gamma_M[k]$ for the multi-channel TF representation is defined, such that if $|\mathbf{MTF}[m,k]|<\gamma_M[k]$ then only ambient noise is considered present. \textcolor{black}{The threshold is calculated by
\begin{equation}
\gamma_M[k]=\min\limits_{m\in T_{pre}} \beta_M[m,k]
\label{eq:Tresh}
\end{equation}
where $T_{pre}$ is the time lapse prior to the appearance of FO. $\beta_M[m,k]$ is the value of $|\mathbf{MTF}[m',k']|$ that is exceeded a certain proportion of occurrence (identified with the parameter $L_{FA}$ which we call false alarm level) within a sliding rectangular window centered around each point $[m,k]$ on the TF plane. This calculation is done directly by ordering the values of $|\mathbf{MTF}[m',k']|$ within the window, from highest to lowest, and taking the value corresponding to $L_{FA}$.} Window overlapping makes the threshold variation with frequency smooth.
Similarly, thresholds are defined for each of the signals, which will then be used in the filtering process. For example, $\gamma_{P^{i}}[k]$ is the threshold for active power flow of branch $i$, calculated from $|\mathbf{TF}_{P^i}[m,k]|$.

\textcolor{black}{The proposed threshold definition could be interpreted as a heuristic extension for non-stationary signal of the thresholds applied on estimates of the power spectral density \cite{Follum2016}, but using the absolute value of the TFR coefficients instead.}
In this work, we use a sliding window of dimensions 0.1 Hz and 25 seconds centered at each point, \textcolor{black}{and $L_{FA}=0.03$.} \textcolor{black}{The simulated data and PMU data used in this paper present an initial time interval of approximately 30 seconds before the appearance of the FOs. For this reason, the temporal width of the sliding window was set in 25 s, but if this calculation were carried out on longer time series, then the most convenient width of the window could be analyzed.} 

\subsection{Detection and Identification of TF Ridges}
In order to find the dominant ridges, the following optimization problem should be solved
\begin{equation}
\max_{C}\sum_{h=1}^H\int_{\mathbb{R}} |MTF(t,\varphi_h(t))|^2 dt 
\label{eq:ridges_opt}
\end{equation}
\textcolor{black}{where the cost function to be maximize takes into account the modulus of the TFR in order to find the set of estimated ridges $C = \{ \varphi_1(t),...,\varphi_H(t)\}$ over which the energy of the spectrogram is maximum.}
We use algorithm 1 to compute an estimate of the ridges with higher energy for a TF representation $\mathbf{MTF}[m,k]$, and the resulting ridges are identified in discrete time index $ \varphi_1[m],...,\varphi_H[m]$. Algorithm 1 is a modified version of the algorithm in \cite{Meignen2017}, the difference is that the search in the TF plane is performed in jumps instead of searching between contiguous time steps, allowing better performance in high-noise conditions. In addition, a frequency-dependent threshold $\gamma_M[k]$ is used, which improves the performance of the algorithm in power systems applications where the ambient noise is colored. Furthermore, the new modified algorithm allows establishing when an oscillatory component begins or ends within the analysis window, a matter not contemplated in the original algorithm.

\begin{algorithm}[h]

 \For{h=1:H}{
 \For{u=1:U}{

  1. Define time interval for initial point
  
  $I_u=[(u-1)L/U,uL/U]$.
  
  2. Find $[q_0,c_u[q_0]]= \argmax\limits_{m\in I_u,k\in [0,N/2-1]} |\mathbf{MTF}[m,k]|$
  
  3. Iterate forward in time from $q=q_0$ 

  \While {$|\mathbf{MTF}[q,c_u[q]]|>\gamma_M[c_u[q]] \And q\leq L$} {

  $[q*,c_u[q*]]= \argmax\limits_{[m,k]\in A_q^+} |\mathbf{MTF}[m,k]|$ 
  
  Linear interpolation of intermediate points
  
  $q=q*$}

   4. Iterate backward in time from $q=q_0$  

  \While {$|\mathbf{MTF}[q,c_u[q]]|>\gamma_M[c_u[q]] \And q \geq 1$} {
  
  $[q*,c_u[q*]]= \argmax\limits_{[m,k]\in A_q^-} |\mathbf{MTF}[m,k]|$ 
  
 Linear interpolation of intermediate points
  
  $q=q*$}

  5. Calculate energy 

  $E_{c_u}=\sum\limits_q  |\mathbf{MTF}[q,c_u[q]]|^2$
  
 } 
 
 6. Adopt the highest energy curve whose length is greater than $L_c$ 
 
 $\mathbf{\varphi_h}[m]=\argmax\limits_{c_u/\text{length} (c_u)>L_c} E_{c_u} $
 
 7. Peel component
 
 $\mathbf{MTF}=\mathbf{MTF} \setminus \bigcup_m[ \mathbf{\varphi_h}[m]-d_1, \mathbf{\varphi_h}[m]+d_1]$

 }
 \caption{Ridge Detection and Identification}
\end{algorithm}
\vspace*{-2mm}

\begin{figure}[h]
\centering
\vspace*{-2mm}
\includegraphics[width = 1\columnwidth]{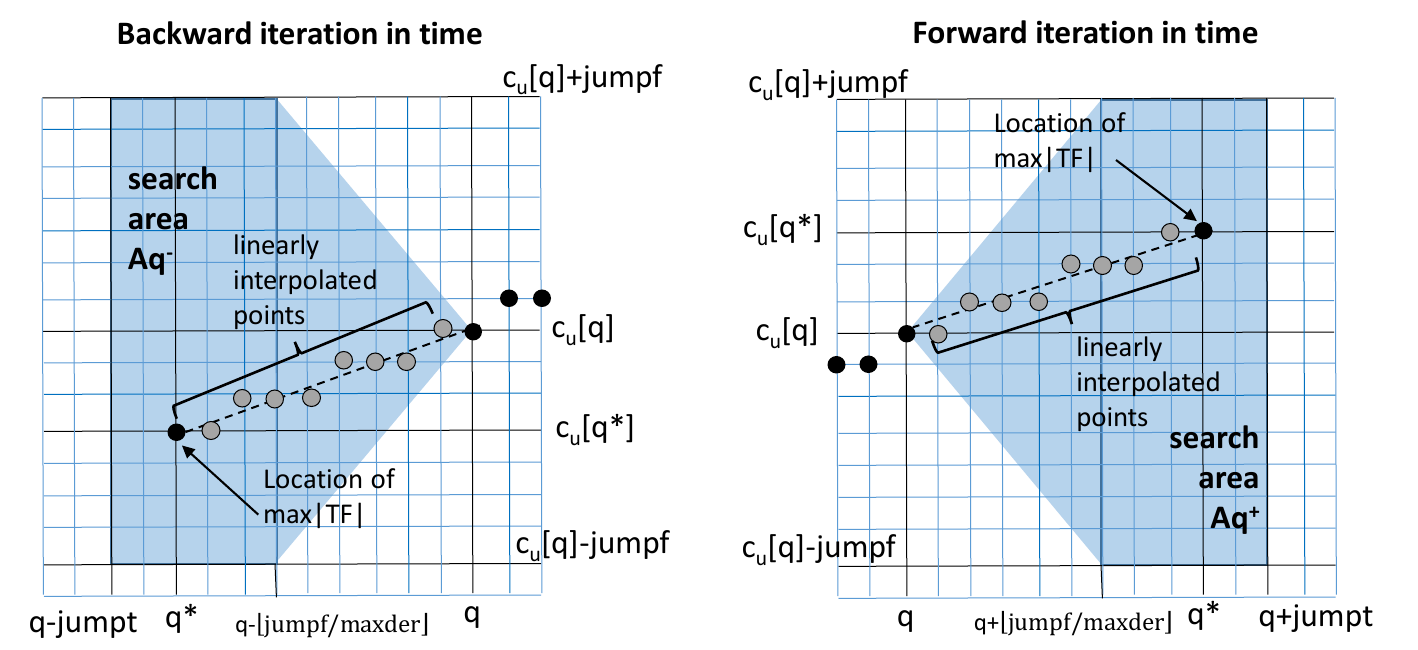}
\vspace*{-4mm}
\caption{Ridge identification algorithm search region}
\vspace*{-2mm}
\label{fig:salto}
\end{figure}

For the estimation of one ridge curve, the algorithm identifies U candidate curves from different initial points (chosen as the maximum of $\mathbf{TF}_f$ in U time intervals).
From that initial point, the algorithm first sweeps the TF plane in a positive direction of time (forward iteration) looking for a local maximum within the region described in Fig. \ref{fig:salto}. The parameters that determine the search areas ($A_q^+$ and $A_q^-$) are $jumpt$ (maximum temporal jump), $jumpf$ (maximum frequency jump) and $maxder$, which is the maximum derivative of frequency to limit frequency modulation of identified curves. The search is carried out while $|\mathbf{MTF}|$ is greater than the threshold $\gamma_M[k]$ beyond which the component is assumed to vanish. Then it performs an analogous iteration from the initial point but in the negative direction of time (backward iteration). We use parameter $jumpt$ corresponding to 2 seconds, $jumpf$ corresponding to 0.03 Hz and $maxder$ corresponding to 0.03 Hz/s. The candidate curve with the highest energy is chosen whose length is greater than a certain duration of time $L_c$ (for example, a minimum of 40 seconds is considered to avoid capturing components that represent damped natural oscillations). Identified curves with duration less than $L_c$ are discarded. Then, the peeling action is carried out setting the coefficients associated with it to zero, in order to continue applying the algorithm to identify another ridge. Set subtraction operation is indicated with $\setminus$ in algorithm 1. $d_1$ is the thickness used to peel the ridge associated with an identified mode. For STFT we choose $d_1=\lfloor3std_{\hat{g}}N/Fs\rceil$, while for FSST or FSST2, due to the fact that the TFR are more concentrated around ridges, we use $d_1=\lfloor std_{\hat{g}}N/Fs)\rceil$, where $\lfloor x \rceil$ is the nearest integer to $x$.

\subsection{Filtering and Reconstruction in Time Domain}
Hard Thresholding (HT) technique, when used for retrieving the component $f_h$ of a MCS, considers in the reconstruction process only the coefficients of the TF representation in the vicinity of $\mathbf{\varphi_h}$, whose magnitude is above a certain threshold \cite{Pham2018} \cite{Laurent2020}. The frequency bands around each ridge used for mode reconstruction $J_{1,h}[m]=[\eta_{1,h}^{-}[m],\eta_{1,h}^{+}[m]]$ are 
\begin{align}
\eta_{1,h}^{-}[m]=\max \Big\{\eta_h^{-}[m], \varphi_h[m]-d_{max} \Big\} \\
\eta_{1,h}^{+}[m]=\min \Big\{\eta_h^{+}[m],\varphi_h[m]+d_{max}\Big\} \label{eq:HTinterval2}
\end{align}

where 
\begin{align}
\eta_h^{-}[m]=\argmax\limits_{k<\varphi_h[m]-d_{min}}\Big\{|\mathbf{TF}_{f}^g[m,k]|<\gamma_f[k]\Big\} \\
\eta_h^{+}[m]=\argmin\limits_{k>\varphi_h[m]+d_{min}} \Big\{|\mathbf{TF}_{f}^g[m,k]|<\gamma_f[k]\Big\} 
\label{eq:HTinterval}
\end{align}

$\gamma_f[k]$ is frequency dependent threshold, $d_{min}$ is the minimum band from which the threshold is checked and $d_{max}$ is the maximum value for the band to avoid overlap of close modes and to establish a limit in low noise signals. The reconstruction in time domain then results (with $g(0)=1$): 
\begin{equation}
\mathbf{f}_h[m]=\sum_{k\in J_{1,h}[m]} \mathbf{TF}_{f}^g[m,k]
\label{eq:reconHT}
\end{equation}

This reconstruction procedure can be applied to STFT, FSST and FSST2.

\section{Numerical Results}
In this section, we present the analysis of different examples applying STFT, FSST or FSST2. First, the proposed methodology is applied to simulated data in WECC179 model. Secondly, the proposed methodology is applied to PMU measurements of a real-world event occurred in ISO-NE \cite{Maslennikov2018}. Finally, we show the results on some representive cases of 2021 IEEE-NASPI OSL Contest \cite{OSL}. The code for the application of STFT, FSST, FSST2 is a modified version of \cite{Laurent2020code}. The choice of parameter $\sigma$ of the window is done by considering the minimal Rényi entropy of the FSST spectrum \cite{Gill2020}. Since the frequency band of electromechanical phenomena in power systems is usually between 0.1 Hz and 2.5 Hz, a parameter $\sigma$ that gives good results for different circumstances is the corresponding to $std_g$ in the range from 2s to 10s. Once defined, the value of $\Delta=3std_{\hat{g}}$ is an indication to verify the condition of well separated modes in frequency.

\begin{figure}[h]
\centering
\includegraphics[width = 0.70\columnwidth]{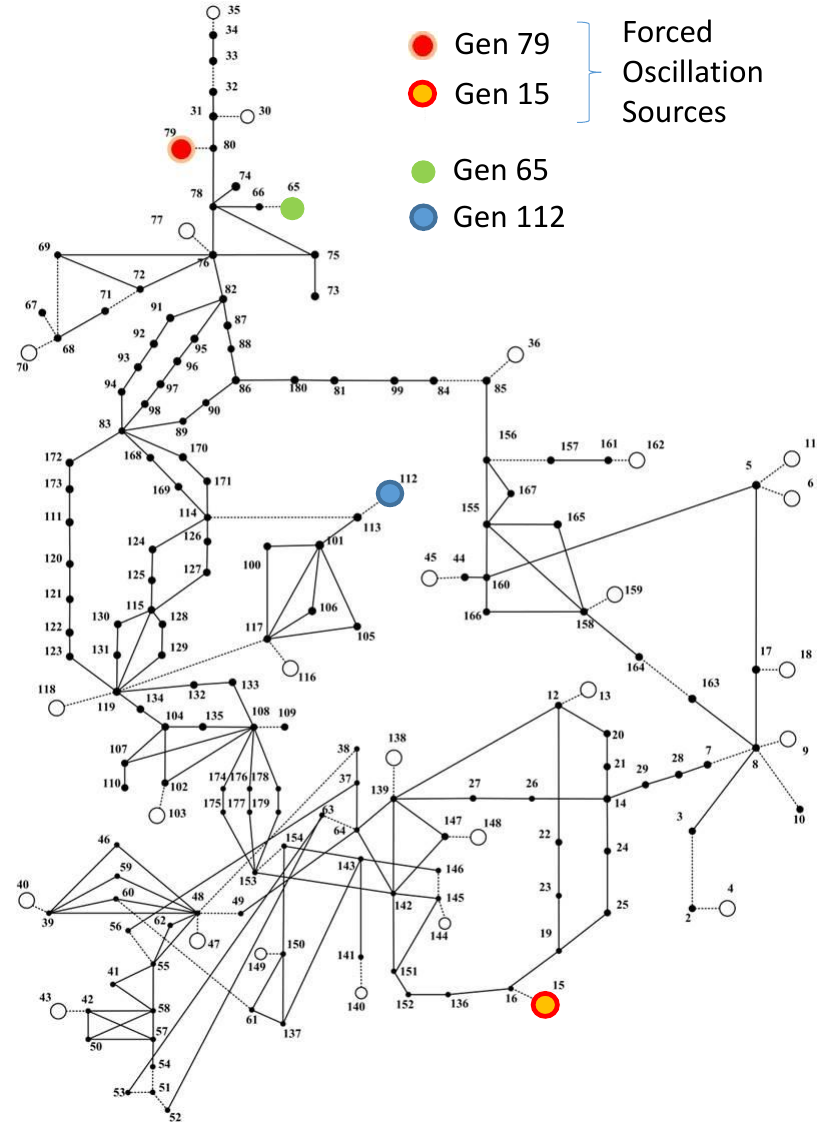}
\vspace*{-2mm}
\caption{WECC 179 bus system. Model data is obtained from \cite{Maslennikov2018}}
\label{fig:WECC179}
\end{figure}

\begin{figure}[h]
\centering
\includegraphics[width = 0.9\columnwidth]{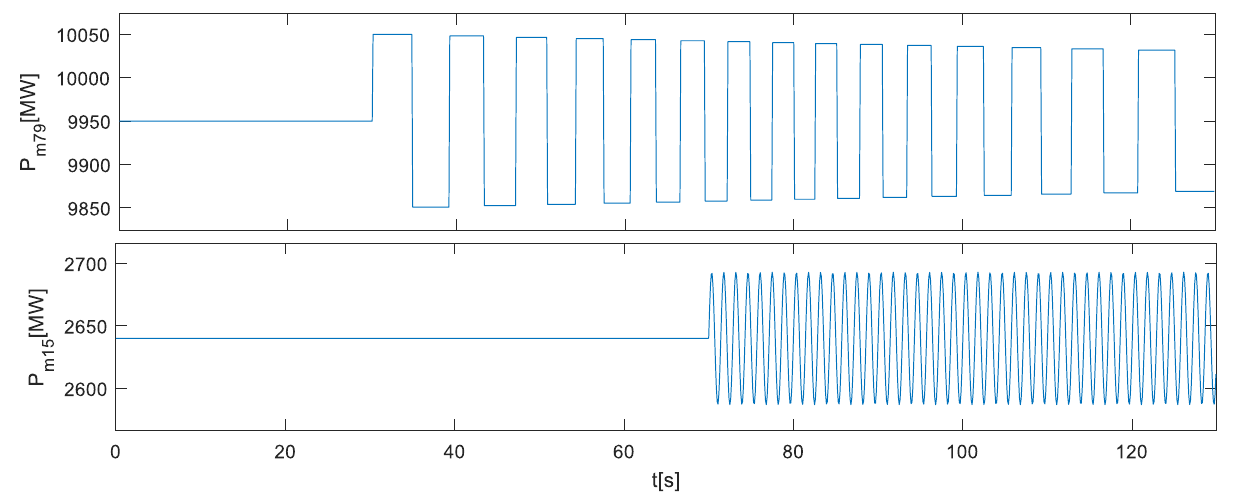}
\vspace*{-3mm}
\caption{Mechanical power of generator 79 and generator 15 (sources)}
\label{fig:Pmec}
\end{figure}

\begin{figure*}[h]
\centering
\setlength\fboxsep{0pt}
\setlength\fboxrule{0.25pt}
\includegraphics[width=1.85\columnwidth]{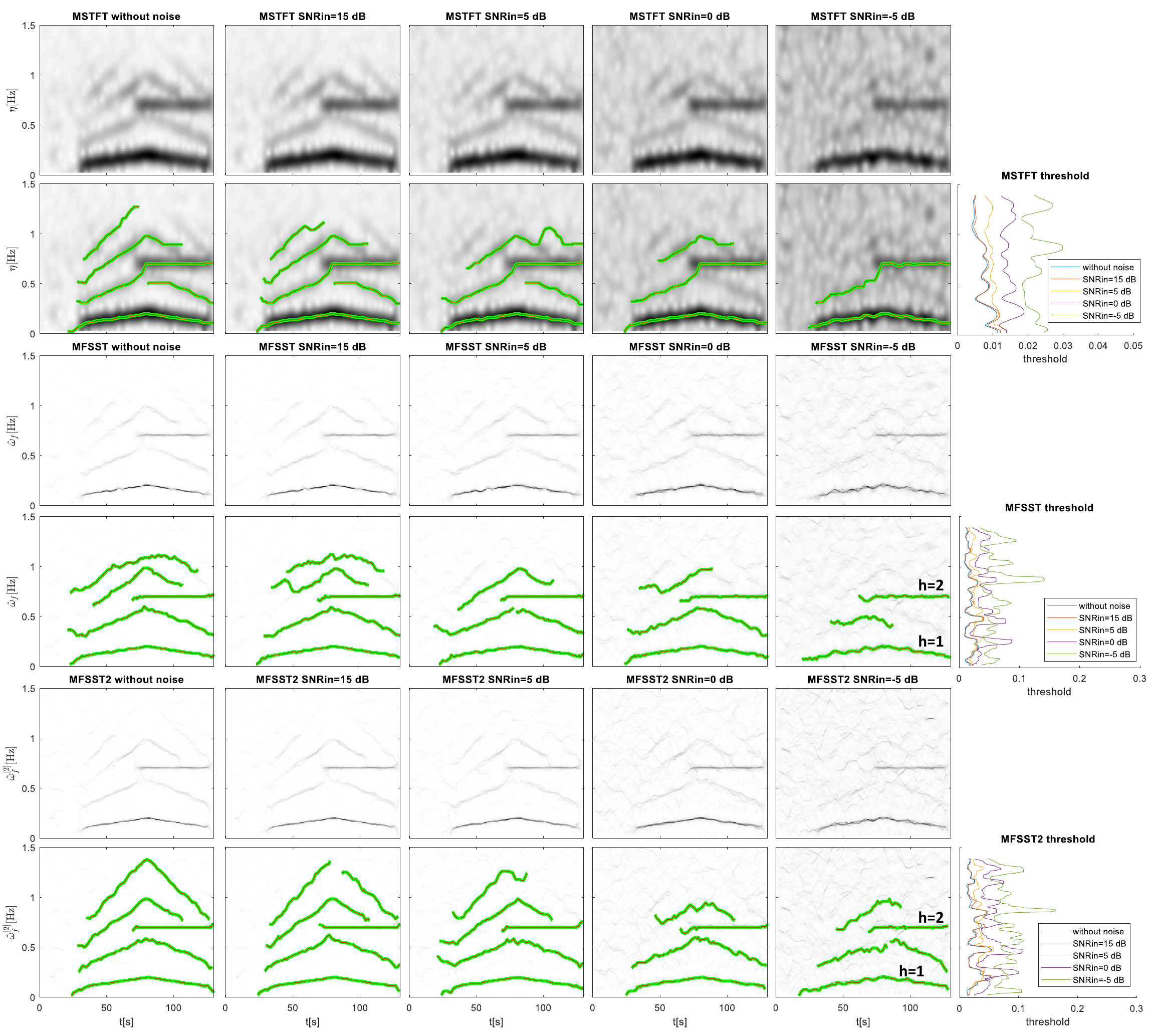}
\vspace*{-3mm}
\caption{Comparison of the application of algorithm 1 for frequency identification on multi-channel TFR based on STFT, FSST and FSST2, with different levels of additive noise. Identified ridges are shown as green curves in respective plots. Simulated Case WECC 179 bus system.}
\label{fig:multiTF_ridges2}
\end{figure*}

\subsection{WECC 179 bus}

The described methodology for the identification of oscillatory components are applied to simulated data from the WECC 179 model, whose one-line diagram is shown in Fig. \ref{fig:WECC179}. A non-stationary mechanical power is applied in generator 79 using a square signal whose fundamental frequency is linearly increased from 0.1 Hz to 0.2 Hz from $t=30$s to $t=80$s, and then linearly reduced to 0.1 Hz in other 50 seconds. This non-sinusoidal input with variable frequency produces a series of variable frequency harmonics that could interact with system natural modes at multiple frequencies. Additionally, a constant frequency disturbance of 0.7 Hz is added to the mechanical power of generator 15 at $t = 70$s (shown in Fig. \ref{fig:Pmec}). White Gaussian noise was introduced in the loads during the simulation in order to generate colored ambient noise in the system magnitudes. Additive white noise with different levels was also added to the simulation results in order to evaluate the noise tolerance of the proposed algorithms.

\subsubsection{Frequency Identification}
We denote by $\alpha(t)$ an additive white Gaussian noise with zero mean and variance $\sigma_\alpha^2$, which is added to the simulated signals. The Signal-to-Noise Ratio (SNR) in dB will be defined by
\begin{equation}
SNRin[dB] =10 \log_{10} (Var(f) / \sigma_\alpha^2)
\label{eq:SNRin}
\end{equation}
where $Var(f)$ is the variance of the noiseless signal $f$ that contains the FO, resulting from simulation. Fig. \ref{fig:multiTF_ridges2} graphically shows the results of the application of the multi-channel methodology using active and reactive power of all the generator in the system  with three TF representations (STFT, FSST and FSST2), and different levels of additive noise. Algorithms are applied to identify a maximum of 7 components in each case, but the identified curves whose length is less than $L_c=40$ seconds are discarded. The ridges resulting from the identification algorithm are shown in green. A pattern of time-varying components is observed. Even for low additive noise levels, ridge identification on STFT presents mode mixing problems, where the ridges include the constant frequency component of 0.7 Hz together with the third harmonic of the square signal. FSST and FSST2 significantly reduce mode mixing problems because the spectrum is much more concentrated. In the case of FSST, it is not possible to identify the higher order harmonics due to their higher frequency modulation, even at low levels of added noise. For example, identified curves from the FSST are correct only up to the fifth harmonic (that has a modulation of 0.01 Hz/s). On the other hand, using FSST2 it is possible to identify the harmonics with the highest frequency modulation. In both (FSST and FSST2), as the addite noise increases, the lower amplitude harmonics are lost. However, even for an SNR of -5dB, it is possible to identify the fundamental frequencies of both sources of oscillation with both FSST and FSST2.

\subsubsection{Filtering}
\begin{figure}[h]
\centering
\includegraphics[width = 0.89\columnwidth]{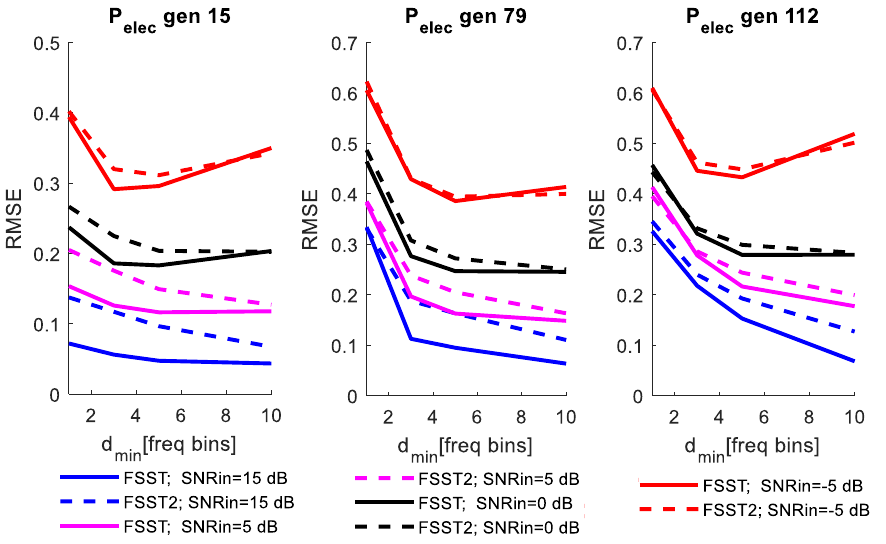}
\vspace*{-2mm}
\caption{Root mean squared error RMSE of fundamental frequency components in electric power for different values of reconstruction band $d_{min}$}
\vspace*{-2mm}
\label{fig:Pelec_dmin}
\end{figure}

In order to compare the reconstruction error of each method, we define a root mean squared error by
\begin{equation}
RMSE=\frac{||\hat{f}-f||_2}{||f||_2}
\label{eq:RMSE}
\end{equation}
where $f$ is the signal without noise and $\hat{f}$ is the reconstructed signal with a specific method, as the sum of the filtered components of each case. First, we analyze the impact of the value of the reconstruction band $d_{min}$ for the FSST and FSST2 methods. Fig. \ref{fig:Pelec_dmin} shows $RMSE$ calculated with the sum of the components $h=1$ and $h=2$, and taking as a reference the sum of the components $h=1$ and $h=2$ of the case without additive noise. $RMSE$ is represented as a function of $d_{min}$ and $SNRin$, and each point is calculated as the average value of 10 realizations of additive noise. 
In these tests we use an upper limit $d_{max}$ corresponding to $3std_{\hat{g}}$, which is never reached because the FSST and FSST2 spectrum are highly concentrated around the ridges.
For low noise levels, it is observed that the RMSE improves by increasing $d_{min}$. However, for high noise levels after a certain value of $d_{min}$ RMSE begins to increase. It has been observed that a value of $d_{min} = 5$ gives a good performance in both low and high noise levels. A value of $d_{min}=5$ is adopted for the following tests, which results in the same order of magnitude of $std_{\hat{g}}$. On the other hand, it is worth mentioning that for the same reconstruction band and noise level, the error obtained by FSST is slightly lower than the error obtained by FSST2.

\begin{figure}[h]
\centering
\vspace*{-2mm}
\includegraphics[width = 0.9\columnwidth]{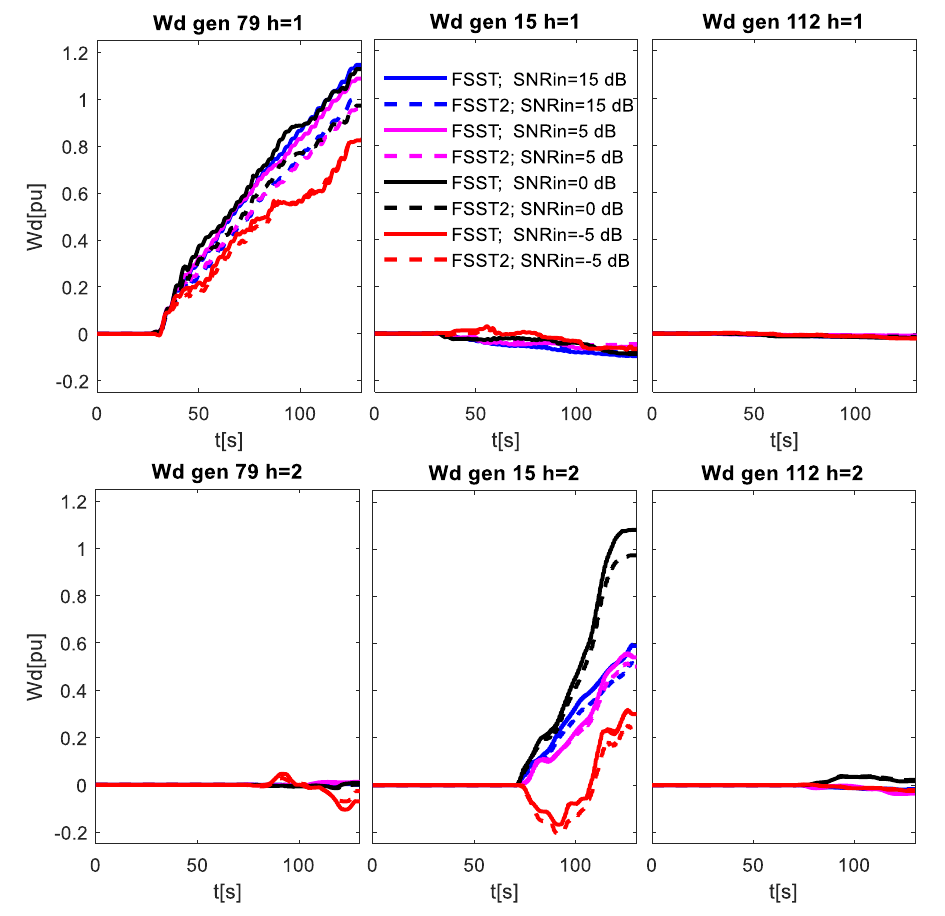}
\vspace*{-2mm}
\caption{Dissipating Energy Flow Wd of different generators calculated from FSST and FSST2 for different levels of added noise.}
\vspace*{-2mm}
\label{fig:Wd_gen}
\end{figure}

\subsubsection{Dissipating Energy Flow}

For the DEF calculation, in addition to filtering the electrical power, it is necessary to perform the decomposition of the reactive power flow signals $Q$, voltage $V$ and voltage angle $\theta$. Fig. \ref{fig:Wd_gen} shows DEF of different generators calculated from FSST and FSST2 for different levels of added noise. Generator 112 is indicated in blue in Fig. \ref{fig:WECC179}, it has been chosen randomly to consider another generator where there is no disturbance. $h = 1$ is the fundamental frequency component of the square signal of the disturbance in generator 79 and $h=2$ corresponds to the 0.7 Hz mode of disturbance in generator 15. In all cases, generator 79 is identified as the source of component $h=1$ and generator 15 as source of component $h=2$. The performance of FSST and FSST2 is very similar in this case.

\subsubsection{Comparison with Conventional DEF Method}
\begin{figure}[t]
\centering
\vspace*{-2mm}
\includegraphics[width = 0.9\columnwidth]{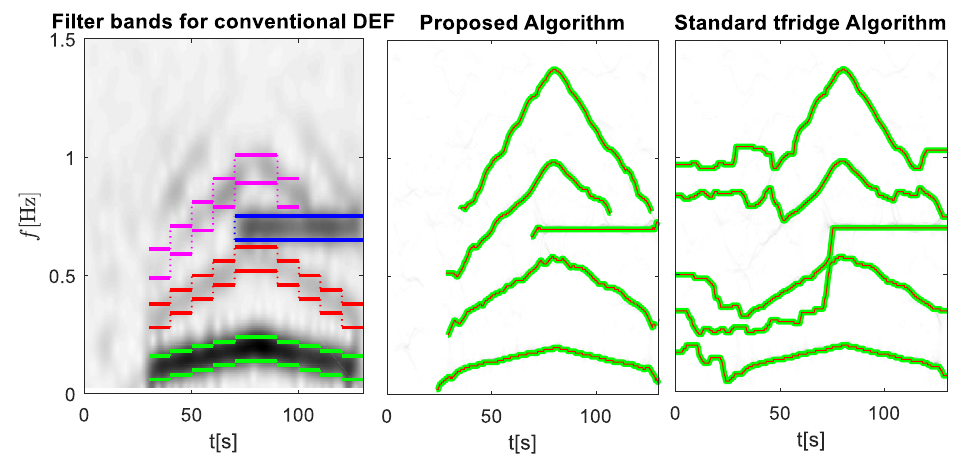}
\vspace*{-3mm}
\caption{Performance comparison of the proposed ridge identification algorithm (center) with respect to multiple band-pass filtering (left) and the standard "tfridge" algorithm (right).}
\vspace*{-1mm}
\label{fig:comp_tfridge}
\end{figure}

\textcolor{black}{An extension of the conventional DEF method for analysis of non-stationary signals could be considered through the application of windowed DFT for the identification of frequencies and the design of multiple band pass filters, splitting the signal into smaller intervals. The challenge in this case would be to first identify the number of oscillatory components from the DFT spectrogram. Then, multiple filter bands should be defined to be able to design the Butterworth filters in each segment of the signal to fit the non-stationary components. 
Assuming that a Gaussian window is used for calculating the windowed DFT and that a multi-channel representation is made using P and Q, then the spectrogram would be the same as the STFT (first row of plots in Fig \ref{fig:multiTF_ridges2}). On this spectrogram, the filter bands should be meticulously defined. For example, Fig. \Ref{fig:comp_tfridge} shows that if the signal is divided into 10-second segments approximately 25 different band-pass filters should be defined, in order to extract the four oscillatory components. It is, at this point, where the proposed methodology has advantages since the ridges detection algorithm automatically identifies the number of non-stationary oscillatory components in the signal. Additionally, with the proposed methodology, it is not necessary to design multiple band pass filters since the reconstruction to the time domain of the signal oscillatory components is carried out directly from the time-frequency representations with a systematic method. When the signal is non-stationary, the filtering approach of the conventional DEF method can be a tedious process that requires special analysis. The proposed methodology provides a systematic and automatic solution to this process. 
On the other hand, the reconstruction error is greater due to the imprecision that occurs when trying to capture the variable frequency by means of a fixed filter band in each signal segment. For example, for components $h$=1 and $h$=2 of the active power of generator 112 with $SNRin=15dB$, the RMSE is 0.8 when multiple Butterworth filters are applied (with the specifications indicated in section II.E). This value is considerably greater than the RMSE of 0.15 and 0.19 that were obtained by applying FSST and FSST2, respectively, with $d_{min}=5$.}

\begin{figure}[t]
\centering
\vspace*{-1mm}
\includegraphics[width = 0.9\columnwidth]{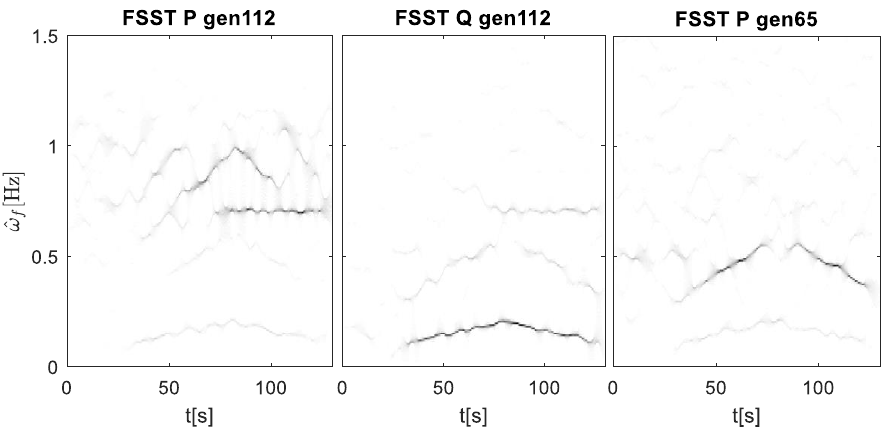}
\vspace*{-3mm}
\caption{FSST of different variables, showing that each one provides information on certain system components. Additive noise SNR=15dB.}
\vspace*{-4mm}
\label{fig:multi_cha}
\end{figure}

\subsubsection{Comparison with our previous work}
\textcolor{black}{In \cite{Gill2020} a standard ridges identification algorithm was used (MatLab function ”tfridge”) for which it is necessary that the oscillatory components of the signal are present throughout the entire analysis window. This is because the algorithm fits each curve for the total duration of the window. Fig. \ref{fig:comp_tfridge} shows performance comparison of the proposed ridge identification algorithm with respect to the standard algorithm used in \cite{Gill2020}. This algorithm applied to this example does not work correctly. 
As it can be observed, mismatches between components are produced, specifically in the instant of time where the FOs begin and end the identified ridges jump between the components. Besides, the identified ridges also capture portion of ambient noise, during the time interval less than 30 seconds, where the FO had not started yet.
On the other hand, in \cite{Gill2020} only the local measurements of the active power was used to identify the FO frequency components. Fig. \ref{fig:multi_cha} shows as example the FSST of active and reactive power of generator 112 (indicated in blue in Fig. \ref{fig:WECC179}) and active power P of generator 65 (indicated in green in Fig. \ref{fig:WECC179}). In particular, P of generator 112 presents higher content in the frequency range between 0.7 and 1.0 Hz, while the reactive power strongly shows the low frequency component (between 0.1 and 0.2 Hz). On the other hand, the active power of generator 65 presents a stronger content in the range of 0.3 to 0.5 Hz. These cases show that the identification of signal components in each channel independently provides only a partial information of frequency components in the system. With the proposed multi-channel methodology, it is possible to identify the components over a single TFR that globally represents the system, as it is shown in Fig. \ref{fig:comp_tfridge}.}

\begin{figure}[t]
\centering
\vspace*{-2mm}
\includegraphics[width = 0.8\columnwidth]{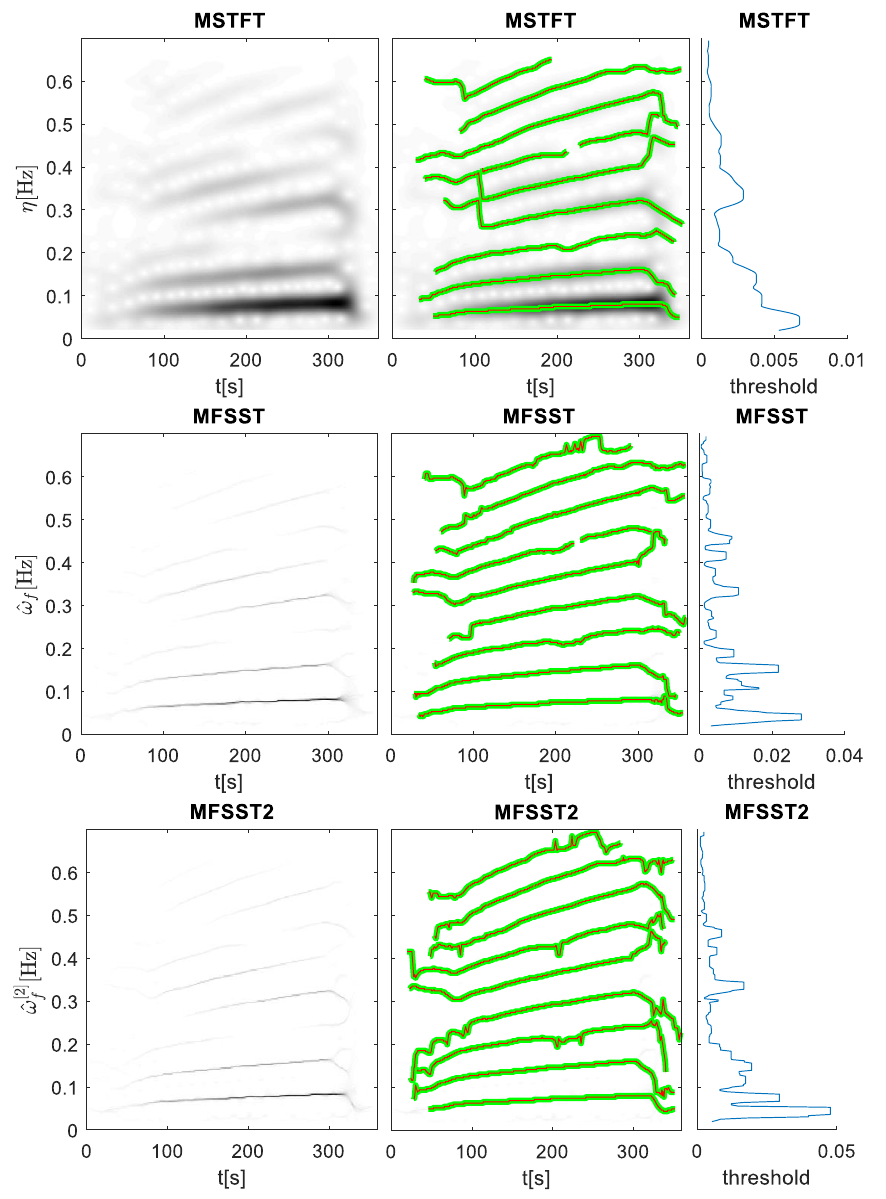}
\vspace*{-2mm}
\caption{Application of multi-channel frequency identification on real PMU data from October 3, 2017 event in ISO-NE }
\vspace*{-3mm}
\label{fig:iso-ne_spec}
\end{figure}

\subsection{ISO-NE October 3, 2017 Event}

Fig. \ref{fig:iso-ne_spec} shows the identified ridges applying the proposed methodology over multi-channel using STFT, FSST and FSST2, alternatively. Multi-channel TF representations are calculated from the measurements of the active and reactive power flows of the 32 monitored branches, whose data were obtained from \cite{Maslennikov2018}. As in the simulated case, FSST and FSST2 have an advantage over STFT for the identification of ridges, avoiding the mixture of components. 
In \cite{Gill2020} the oscillatory components of this event were identified over the FSST spectrum of only one representative signal and in a time window, previously determined where oscillation was clearly present with the naked eye. Additionally, it had been necessary to specify in advance how many oscillatory components are present in the signal. The new approach, does not need to take any prior consideration on the signal (due to multi-channel analysis), nor about the time instance where the FO is presence or the number of components. Once the ridges have been identified in the TF plane, active power flows $P$, reactive power flow signals $Q$, voltage $V$ and voltage angle $\theta$ can be decomposed, in order to calculate DEF to trace the source of oscillation. For space reasons, reconstruction to time domain and DEF results are not presented for this case.


\begin{figure}[h]
\centering
\vspace*{-2mm}
\includegraphics[width = 0.9\columnwidth]{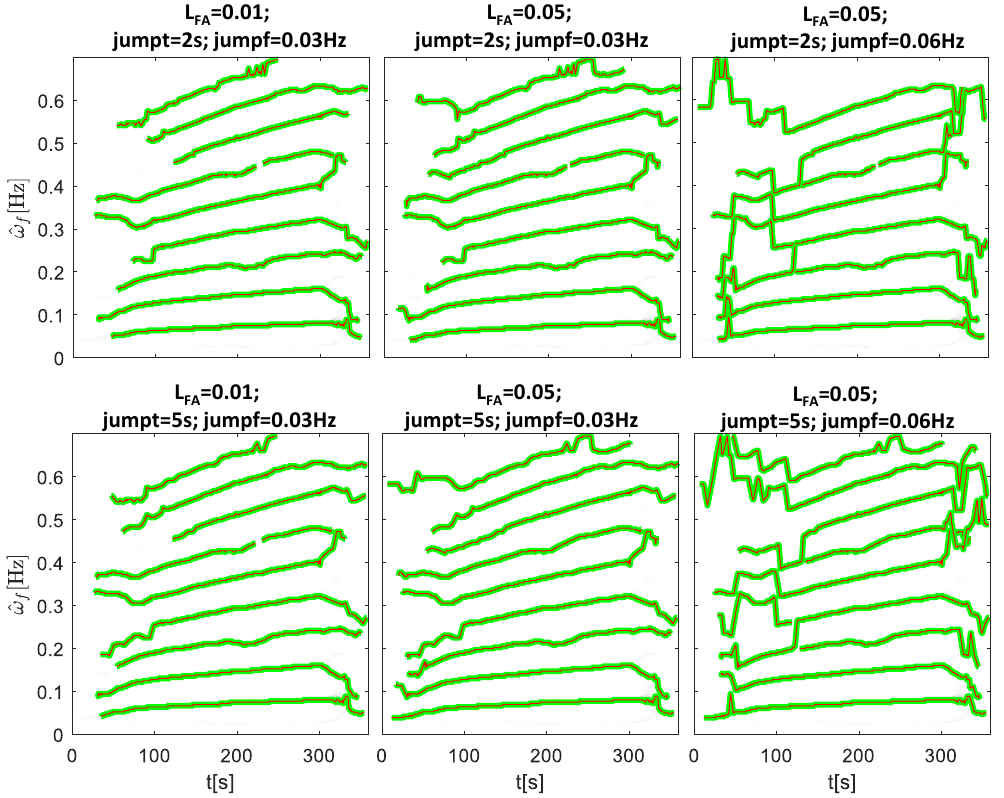}
\vspace*{-2mm}
\caption{Effect of parameter variation on ridge identification algorithm. MFSST of PMU data from October 3, 2017 event in ISO-NE}
\label{fig:param}
\end{figure}

\subsubsection{Parameter selection}
\textcolor{black}{Ridge identification performed in Fig. \ref{fig:iso-ne_spec} was done with the following base parameters: $L_{FA}$=0.03, $jumpt$=2s, $jumpf$=0.3Hz. In Fig. \ref{fig:param} we show the effect of parameter variation on ridge identification algorithm performance.
In particular, the effect of $L_{FA}$ goes hand in hand with the width of the search area ($jumpt$). Admitting a larger $L_{FA}$ implies that weaker magnitudes could be detected. Also, a larger value of $jumpt$ implies that the search algorithm can make longer jumps in time. With both parameters, the sensitivity of the ridge detection and identification algorithm can be increased, but with the risk that the detected ridges extend more than necessary, capturing part of the ambient noise.
On the other hand, the frequency height of the search area ($jumpf$) must be less than the separation of the oscillatory components. For example, in this case when $jumpf$ is greater than 0.06Hz, component mixing occurs, since it is very close to the minimum fundamental frequency of the FO.}

\begin{figure}[t]
\centering
\vspace*{-2mm}
\includegraphics[width = 1\columnwidth]{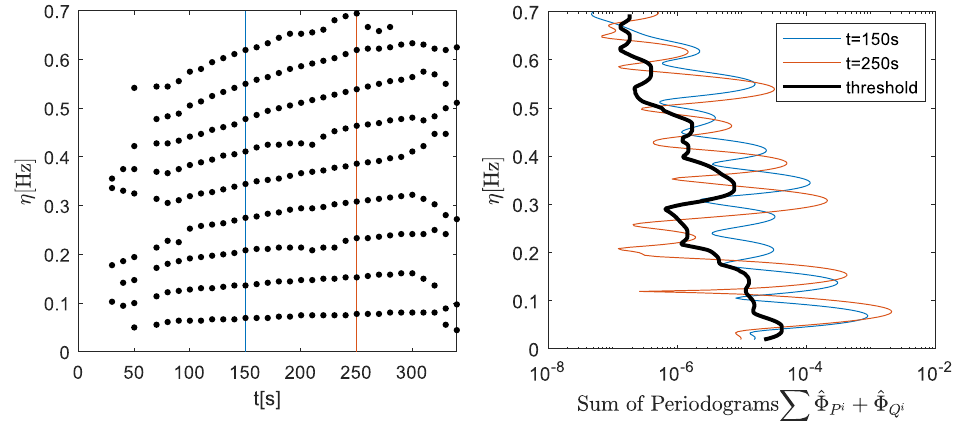}
\vspace*{-3mm}
\caption{FO Frequency detection using periodogram-based detector. PMU data from October 3, 2017 event in ISO-NE}
\vspace*{-3mm}
\label{fig:detec_Periodogram_v1}
\end{figure}
\subsubsection{Comparison with other FO detection methods}
\textcolor{black}{Fig. \ref{fig:detec_Periodogram_v1} shows the result of applying a detection method based on the increase in signal energy (similar to \cite{Follum2016b}). 
In this case, for the calculation of the periodogram we apply a Gaussian window identical to the one used to calculate the STFT, FSTT and FSST2, with a standard deviation of $std_g=10s$. The magnitude we use to determine the presence of FO is the sum of the windowed periodograms of the active power $\hat{\Phi}_{P^i}$ and reactive power $\hat{\Phi}_{Q^i}$ of the monitored branches.
In this case, the frequency-dependent threshold is calculated in the same way as in the proposed methodology, using the results of the sum of windowed periodogram instead of the TFR coefficients. When the sum of the periodograms at a given instant exceeds the threshold, the frequency of the FO is estimated as the value where the peak of this magnitude occurs.
For example, to the right of Fig. \ref{fig:detec_Periodogram_v1} the value of the sum of periodograms for two instants of time $t$=150s and $t$=250s is shown. The peak values of these magnitudes above the threshold correspond to the dots in the TF diagram on the left.}
\textcolor{black}{The detection method manages to correctly detect the oscillatory frequency components at a given instant, but it does not cover the association of the different peaks to a ridge curve in TF plane for each oscillatory component. Another limitation of this method is that it does not contemplate the transformation of oscillatory components to the time domain, that could be necessary for the subsequent application of an OSL method. In this case, some complementary filtering technique would be required. The method proposed in this paper overcomes these two limitations, since it allows to identify the ridges of each oscillatory component in the TF plane and also allows the reconstruction of the components in the time domain.}

\begin{figure}[t]
\centering
\vspace*{-2mm}
\includegraphics[width = 0.8\columnwidth]{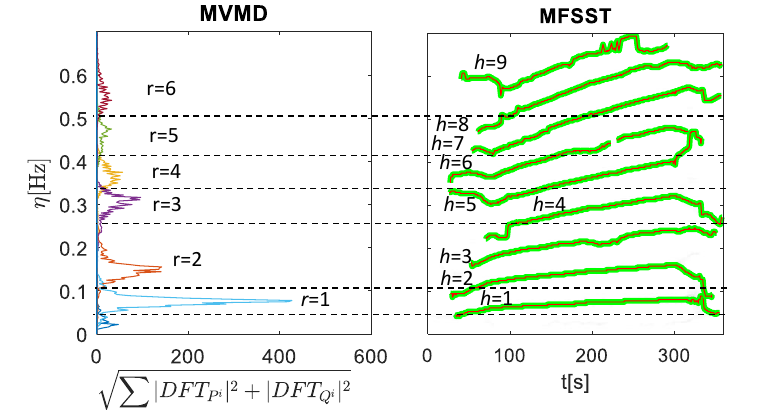}
\vspace*{-2mm}
\caption{Comparison with Multivariate Variational Mode Decomposition (MVMD). PMU data from October 3, 2017 event in ISO-NE}
\vspace*{-3mm}
\label{fig:MVMD_caso2}
\end{figure}
\subsubsection{Comparison with other data-driven signal processing methods (MVMD)}
\textcolor{black}{We compare the performance with respect to a multivariable version of VMD (MVMD) method \cite{Rehman2019}. We use the code of MVMD available in \cite{MVMDcode}. MVMD method decomposes the signals into modulated oscillations with limited bandwidth across a center frequency. Fig. \ref{fig:MVMD_caso2} shows the DFT spectra of the components $r=1,...,6$ resulting from applying MVMD, where the representative frequency bandwidth of each component can be observed. The first component $r$=1 practically coincides with the component $h$=1 identified with MFSST, and it is the fundamental component of the FO that presents the lowest frequency modulation. The second component $r$=2 spans a frequency width that covers the second and third harmonics $h$=2 and $h$=3. In the case of $r$=3, a mixture of components is already beginning to be produced, since it encompasses the fourth harmonic $h$=4 and also part of the fifth harmonic $h$=5.
As the order of the harmonics increases, the frequency modulation increases and the fixed frequency band approach of MVMD method fails to differentiate the variable frequency components from one another. For example, the $r$=4 component of MVMD comprises a frequency band that covers the fifth harmonic $h$=5 and the sixth harmonic $h$=6, without the possibility of distinguishing them.
It should be noted that, if the frequency content of the oscillatory components had remained bounded and separated throughout the entire analysis period, then MVMD would have been able to carry out the decomposition correctly without mixing components.}

\subsection{IEEE-NASPI Oscillation Source Location Contest}

The proposed methodology was applied in the IEEE-NASPI Oscillation Source Location (OSL) Contest \cite{OSL}. A total of 13 cases were studied with the following characteristics. Synthetic PMU measurements of bus voltage and branch current phasors from multiple locations of the test system were provided. There was a 30 seconds leading window before the event and 60 seconds time window after that, total 90 seconds of data. White noise was added to the load during simulation to mimic random load fluctuations. EPRI's PMU Emulator was used to process the simulation results to mimic PMU device performance, a mix of P Class and M Class PMUs were used. EPRI's Synchrophasor Data Conditioning Tool was used to process the synthetic PMU data to introduce data quality problems \cite{OSL}. From a participation of 21 teams, our Team reached third place \cite{OSL}. \textcolor{black}{Below, we present some of the most representative cases.}  

\subsubsection{Case 5 OSL Contest}
\begin{figure}[t]
\centering
\vspace*{-2mm}
\includegraphics[width = 0.9\columnwidth]{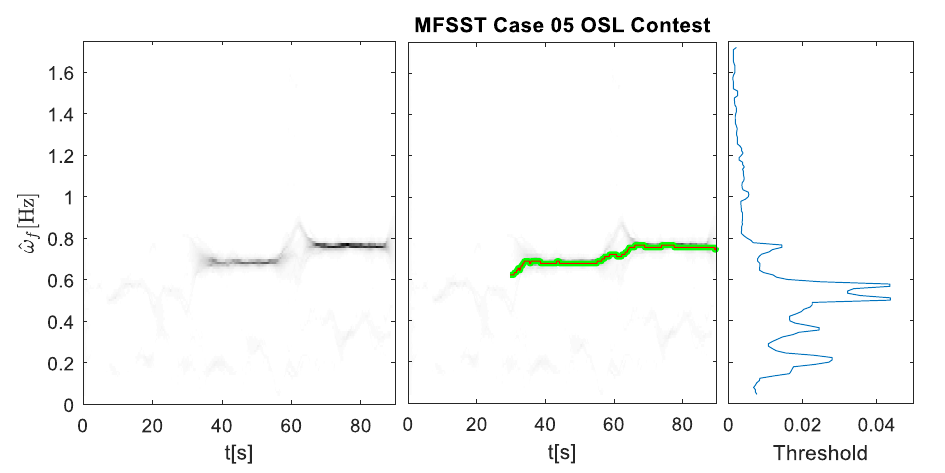}
\vspace*{-2mm}
\caption{Case 5 MFSST, ridge identification and threshold }
\vspace*{-1mm}
\label{fig:Caso05_MSST}
\end{figure}

\begin{figure}[t]
\centering
\vspace*{-1mm}
\includegraphics[width = 0.9\columnwidth]{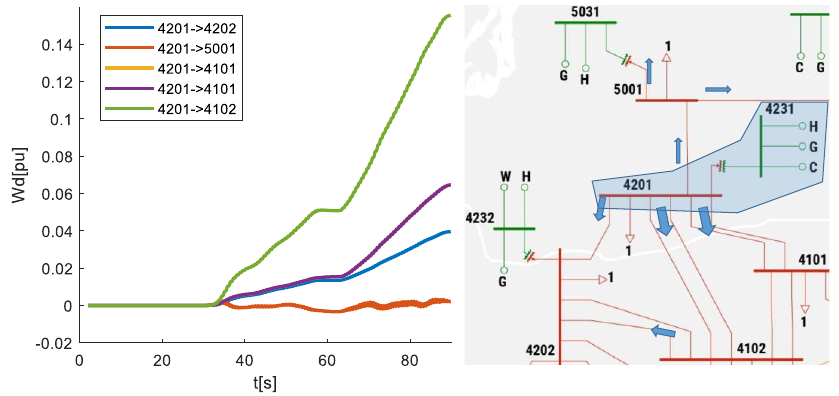}
\vspace*{-2mm}
\caption{Case 5 DEF of each oscillatory component and source localization.}
\vspace*{-2mm}
\label{fig:Caso05_Wd}
\end{figure}

\textcolor{black}{This case presents a variable frequency of FO. Forcing frequency is 0.68Hz before t=58s, 0.76Hz after t=61s, and is transitioning in the 3-sec interval. System has natural modes at: 0.614Hz, 0.708Hz, 0.741Hz and 0.78Hz \cite{OSL}. Fig. \ref{fig:Caso05_MSST} shows the resulting multi-channel TFR using FSST, the threshold calculation and the result of applying the ridge identification algorithm. The algorithm can successfully track frequency variations. 
The calculation of the DEF from the filtered variables using the estimated ridge is shown in Fig. \ref{fig:Caso05_Wd}. Blue arrows indicate the sign of the rate of change of the DEF. In this case, it can be concluded that the source is bus 4231.}

\begin{figure}[t]
\centering
\vspace*{-2mm}
\includegraphics[width = 0.9\columnwidth]{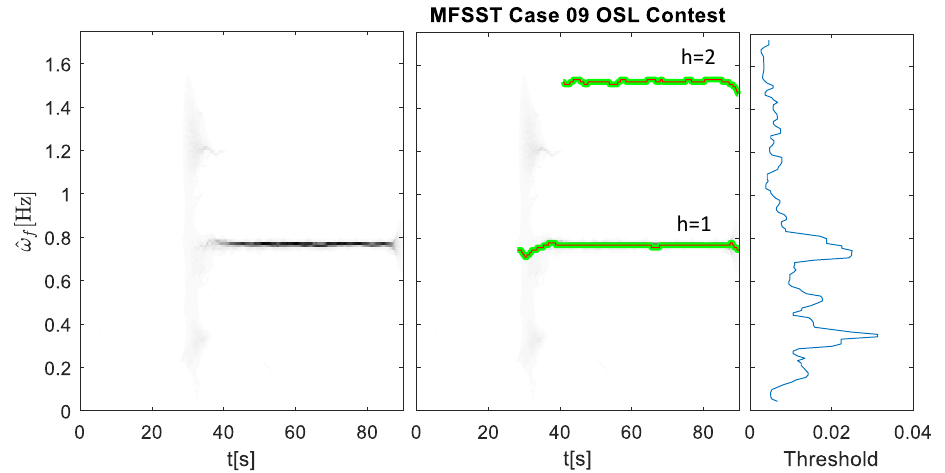}
\vspace*{-2mm}
\caption{Case 9 MFSST, ridge identification and threshold }
\vspace*{-1mm}
\label{fig:Caso09_MSST}
\end{figure}

\begin{figure}[t]
\centering
\vspace*{-2mm}
\includegraphics[width = 0.9\columnwidth]{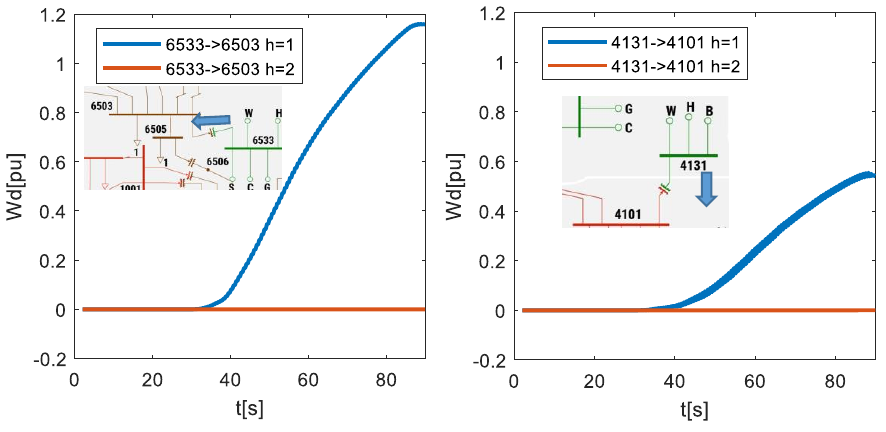}
\vspace*{-2mm}
\caption{Case 9 DEF of each oscillatory component and source localization.}
\vspace*{-3mm}
\label{fig:Caso09_Wd}
\end{figure}

\begin{figure}[b]
\centering
\vspace*{-4mm}
\includegraphics[width = 0.9\columnwidth]{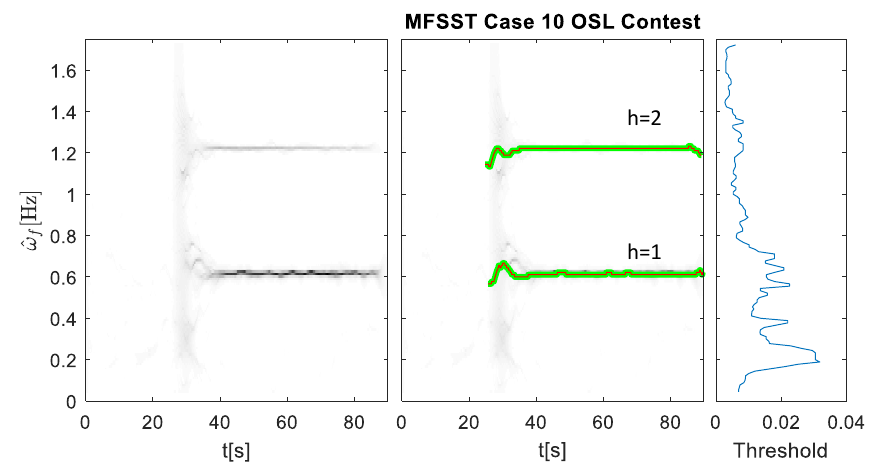}
\vspace*{-2mm}
\caption{Case 10 MFSST, ridge identification and threshold }
\vspace*{-2mm}
\label{fig:Caso10_MSST}
\end{figure}

\begin{figure}[b]
\centering
\vspace*{-2mm}
\includegraphics[width = 0.9\columnwidth]{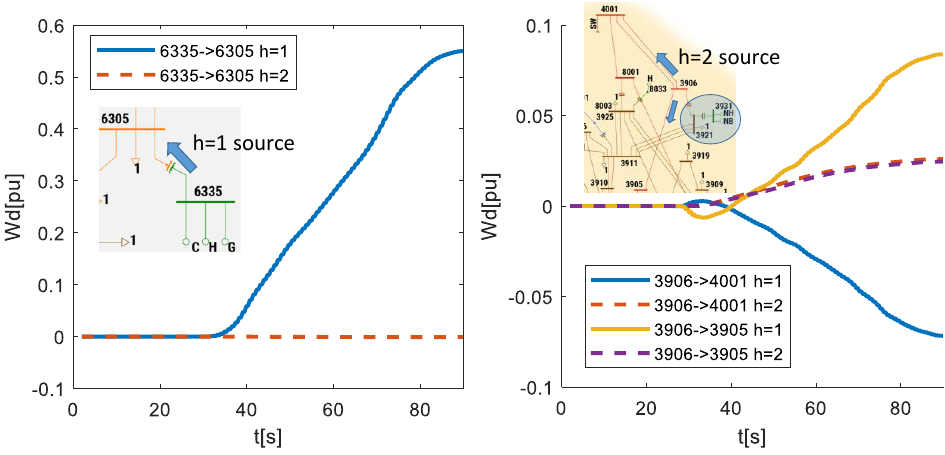}
\vspace*{-2mm}
\caption{Case 10 DEF of each oscillatory component and source localization.}
\vspace*{-2mm}
\label{fig:Caso10_Wd}
\end{figure}

\subsubsection{Case 9 OSL Contest}
\textcolor{black}{The FO (at bus 6533) with a frequency of 0.762 Hz resonates with a natural mode whose damping is reduced by adjusting PSS gain (Ks=-2) in generator 4131 H creating negative contribution into damping from that generator. Maximum oscillation amplitude in MW flow is not at the source \cite{OSL}. Fig.  \ref{fig:Caso09_MSST} shows that in addition to the main frequency component of the FO ($h$=1), a second component ($h$=2) is also detected with the proposed methodology. From the DEF calculation (shown in Fig. \ref{fig:Caso09_Wd}) it can be concluded that the negative damping sources for the main component $h$=1 are the 6533 and 4131. The DEF for the component $h$=2 (whose frequency is twice that component $h$=1) is practically negligible, therefore this component is associated with a low magnitude second harmonic of the main component $h$=1.}

\subsubsection{Case 10 OSL Contest}
\textcolor{black}{There are two forced oscillations, each resonates with a natural mode. Maximum oscillation amplitude in MW flow is not at the source \cite{OSL}. Fig. \ref{fig:Caso10_MSST} shows that two components are identified with the proposed method: 0.61 Hz ($h$=1) and 1.22 Hz ($h$=2). Since in this case the frequency of component $h$=2 is twice that of the component $h$=1, in principle one would be tempted to assume that $h$=2 corresponds to the second harmonic of $h$=1. Nevertheless, DEF calculation in this case indicates that they are two independent components: bus 6335 is the source of $h$=1 and 3931 is the source of $h$=2, as illustrated in Fig. \ref{fig:Caso10_Wd}.}

\section{Conclusions}

\textcolor{black}{Most of the methods used for detection, identification and filtering of oscillatory components of FO tend to deal with the problems separately and generally assume that the FO frequency is stationary.}
\textcolor{black}{The main contribution of this paper is to develop a systematic methodology that jointly performs detection, identification and filtering from multi-channel TFR to retrieve the oscillatory components of non-stationary power system FOs.}
\textcolor{black}{In particular, we develop a ridge detection and identification algorithm, and a hard-thresholding reconstruction to time domain of FO oscillatory components based on a spectrum-dependent threshold.}

\textcolor{black}{The proposed methodology is an improvement over our previous work \cite{Gill2020}. In first place, we extend the methodology for any invertible TF representation, not only FSST. Additionally, in \cite{Gill2020} the frequency identification was carried out locally in each generator or branch, without being able to take advantage of the simultaneity characteristics provided by the PMU measurements at different locations. With the new methodology, the use of a multi-channel TFR allows to have all the frequencies present in the entire system. Furthermore, ridge identification algorithm used in \cite{Gill2020} was only limited to cases where the oscillatory component was present in the entire analysis window, without considering the possibility that an oscillatory component of the signal may begin or disappear during the analysis window. Besides, in \cite{Gill2020}, it was also necessary to specify in advance the number of oscillatory components to be extracted. In this case, the new algorithm can detect when an oscillatory component starts or stops and automatically identifies the number of oscillatory components.}

We study the performance on different TF approaches: short-time Fourier transform (STFT), STFT-based synchrosqueezing transform (FSST) and second order FSST (FSST2). It is shown that due to the sharper and more concentrated spectrum around ridges, FSST and FSST2 present a clear advantage over STFT regarding the effectiveness of the ridge identification algorithm, being able to discriminate more accurately the oscillatory components. \textcolor{black}{On the other hand, FSST2 method performs better than FSST for non-stationary oscillations with high frequency modulations. However, if the frequency modulation of the FO is low enough (less than 0.01 Hz/s), then FSST and FSST2 show similar performance. Available PMU measurements of actual power systems FO have shown that high frequency modulations greater than 0.01 Hz/s do not occur normally. Therefore, taking into account that the computational time of the FSST is approximately half that for FSST2, the use of FSST gives a good trade-off between resolution and computational demand.}
Besides, the proposed methodology could be applied with other invertible TFRs such as CWT-based SST \cite{DAUBECHIES2011243} or multisynchrosqueezing transform \cite{Yu2019}. Finally, the effectiveness of the proposed methodology combined with DEF method was applied in simulated data, real-world PMU data and in the cases of the IEEE-NASPI OSL Contest 2021.
\textcolor{black}{As a future work, we propose to carry out a statistical study on the probability of false alarm and the detection probability of the proposed ridge detection algorithm based on the sensitive parameters and the TFR used.}

\bibliographystyle{IEEEtran}

\bibliography{biblio}

\end{document}